\documentclass{article}

\usepackage{arxiv}
\usepackage{fontenc}
\usepackage[utf8]{inputenc}
\usepackage{layout}
\usepackage{xcolor}
\usepackage{siunitx}
\usepackage{comment}
\usepackage{amsmath,amsfonts,amssymb}
\usepackage{booktabs}
\usepackage{multirow}
\usepackage{pdflscape}
\usepackage{bigfoot}
\usepackage{authblk}
\usepackage{etoolbox}
\usepackage{url}            
\usepackage{nicefrac}       
\usepackage{microtype}      
\usepackage{lipsum}         
\usepackage{graphicx}
\usepackage[sorting=none, isbn=false]{biblatex}
\usepackage{doi}
\usepackage{subcaption}
\usepackage{hyperref}
\usepackage{cleveref}       
\hypersetup{
    colorlinks=true,
    linkcolor=blue,
    filecolor=magenta,      
    urlcolor=cyan,
    pdftitle={High-Fidelity-Numerical-Modeling},
    pdfpagemode=FullScreen,
    pdfauthor={Jacopo Bonari, Francesca Marsili, Max von Danwitz, Alexander Popp}, 
    pdfkeywords={structural health monitoring, sensitivity analysis, high-fidelity numerical modeling}
}

\DeclareNewFootnote{AAffil}[arabic]
\DeclareNewFootnote{ANote}[fnsymbol]

\makeatletter
\patchcmd\maketitle{\def\@makefnmark{\rlap{\@textsuperscript{\normalfont\@thefnmark}}}}{}{}{}
\makeatother

\makeatletter
\def\thanksAAffil#1{
  \footnotemarkAAffil\protected@xdef\@thanks{\@thanks%
        \protect\footnotetextAAffil[\the \c@footnoteAAffil]{#1}}%
}
\def\thanksANote#1{%
  \footnotemarkANote%
  \protected@xdef\@thanks{\@thanks%
        \protect\footnotetextANote[\the \c@footnoteANote]{#1}}%
}
\makeatother

\title{Physics-Informed Sensitivity Analysis for Enhanced Structural Health Assessment: Test-Case for a Mixed Steel-Concrete Bridge}

\newbox{\orcid}\sbox{\orcid}{\includegraphics[scale=0.06]{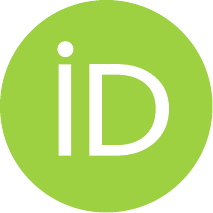}}

\author[1]
{
	\href{https://orcid.org/0000-0001-8435-6466}{\usebox{\orcid}\hspace{1mm}Jacopo~Bonari\thanks{\texttt{jacopo.bonari@dlr.de}}}%
}
\author[2]
{
	\href{https://orcid.org/0000-0002-0098-3428}{\usebox{\orcid}\hspace{1mm}Francesca~Marsili}
}
\author[1]
{
	\href{https://orcid.org/0000-0002-2814-0027}{\usebox{\orcid}\hspace{1mm}Max~von~Danwitz}
}
\author[1,3]
{
	\href{https://orcid.org/0000-0002-8820-466X}{\usebox{\orcid}\hspace{1mm}Alexander~Popp}
}

\affil[1]
{
German Aerospace Center (DLR), Institute for the Protection of Terrestrial Infrastructures, \newline 53757 Sankt Augustin, Germany,
\href{https://www.dlr.de/en/pi/}{www.dlr.de/pi}
}
\affil[2]
{
Chair of Engineering Materials and Building Preservation, Helmut-Schmidt-University, \newline 22043 Hamburg, Germany,
\href{https://www.hsu-hh.de/kwb/en/}{www.hsu-hh.de/kwb/en}
}

\affil[3]
{
Institute for Mathematics and Computer-Based Simulation (IMCS), University of the Bundeswehr Munich, \newline 85577 Neubiberg, Germany,
\href{https://www.unibw.de/imcs-en/}{www.unibw.de/imcs-en}
}

\newcommand{\gr}[1]{\textcolor{gray}{#1}}
\newcommand{\bx}{\boldsymbol{x}}
\newcommand{\bX}{\boldsymbol{X}}
\newcommand{\bY}{\boldsymbol{Y}}
\newcommand{\domain}{\mathcal{D}}
\newcommand{\bal}{\boldsymbol{\alpha}}
\newcommand{\bu}{\boldsymbol{u}}
\newcommand{\Ec}{E_\mathrm{c}}
\newcommand{\db}{d_\mathrm{b}}
\newcommand{\keq}{k_\mathrm{eq}}
\newcommand{\he}{\mathrm{He}}
\newcommand{\bxi}{\boldsymbol{\xi}}

\bibliography{bibliography.bib}

\graphicspath{{figures/}}

\begin{document}

\maketitle

\begin{abstract}
Bridges are vital components of transportation systems, that serve as essential links ensuring the safe and efficient movement of people, goods, and emergency responders, especially during crises. With aging infrastructures, increasing traffic volumes and loads, and the intensifying impacts of extreme weather events due to climate change, the development of effective physics-informed structural health monitoring (SHM) frameworks has become critically important, more so when combined with sensitivity analysis (SA), which identifies the most influential structural parameters in the bridge’s response. To support this, a high-fidelity, physics-based numerical model of a full-scale, two-span, mixed steel-concrete test bridge has been developed. This model serves as a virtual replica of a real structure located at the University of the Bundeswehr Munich. The numerical model is used as a complementary tool to improve prognostic capabilities and quantify uncertainty. A SA study is conducted to evaluate the structure’s response under various mechanical conditions. Assessing these operational variations' effects on structural behavior forms part of an integrated, systematic evaluation framework aimed at combining SHM and SA.

\keywords{structural health monitoring \and sensitivity analysis \and high-fidelity numerical modeling.}
\end{abstract}

\section{Introduction}
Civil infrastructure systems, particularly bridges, represent critical nodes in transportation networks, ensuring the safe and efficient movement of people, goods, and emergency services. The structural integrity of these assets faces growing challenges due to aging materials, increasing traffic loads and volumes, and climate change-induced extreme weather events~\cite{mondoro:2018}. These factors stress the urgent need for advanced structural health monitoring (SHM) frameworks capable of providing reliable, physics-informed assessment of structural performance under uncertainty~\cite{radulescu:2024}.

In this context, high-fidelity, physics-informed finite element (FE) models, either employed alone or coupled with real-time experimental data as to form a digital twin (DT), or a digital shadow~\cite{brucherseifer:2021}, play a crucial role in providing accurate responses and assistance to informed decision making processes. Such models offer the desired level of accuracy in simulating structural behavior, provided they are properly calibrated and validated against experimental data~\cite{schlune:2009}.

The current study describes a sensitivity analysis (SA) framework based on a detailed FE model of a full-scale, two-span steel-concrete composite bridge located at the University of the Bundeswehr Munich~\cite{jaelani:2023,hansen:2023}. The physics-based model serves as a virtual replica of the real structure, enabling an accurate prediction of its dynamic response under environmental conditions. Here, the predictive capability of the numerical model is extended by considering uncertainties in specific input parameters, namely material properties and geometry. To address this problem, a SA framework that evaluates how variations in key structural parameters affect the model output is derived.

By quantifying the sensitivity of the bridge's dynamic response to these parameters, more informed and accurate design of monitoring campaigns can be established, e.g., highlighting correlations with safety related variables, improving the sensors network, and focusing the attention on the quantities to which the model is more sensitive~\cite{resta:2024}. According to these premises, SA establishes a sound and rigorous foundation for a model-based decision-making process that meets expected levels of physical accuracy.


To quantify the relative importance of the uncertain inputs on the model output, an \emph{ANOVA} (ANalysis Of VAriance) technique is employed, that allows to derive the variance of the model output as a sum of contributions from each input variable or, in other terms, to quantify the fraction of the output variance that is determined by the variance of each input. These contributions go under the name of Sobol' indices~\cite{sobol:2001}. To evaluate them, the generalized Polynomial Chaos Expansion (gPCE) method~\cite{xiu:2010} is leveraged, which provides a powerful framework for SA. As a matter of fact, gPCE allows to compute Sobol' indices analitically, thus providing a remarkable advantage over different approaches~\cite{saltelli:2010}.

The resulting insights enable a more comprehensive understanding of the bridge's structural behavior under variation of the selected input parameters, contributing to the development of more robust, physics-informed DT system to support proactive maintenance and risk management strategies in the field of critical infrastructures protection~\cite{danwitz:2023,torzoni:2024}.


\begin{figure}
\includegraphics[width=\textwidth]{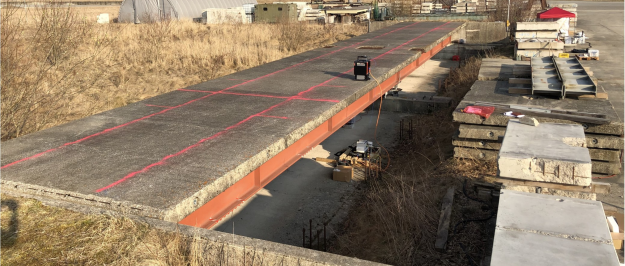}
\caption{Overview of the test bridge, with top deck concrete slabs clearly visible, together with part of the southern principal HEB1000 girder.}\label{fig:global_view}
\end{figure}

\section{High-Fidelity Numerical Model}\label{sec:model}
In order to provide a sound test-case for the development of effective physics-informed SHM models and practices, a high-fidelity FE model of a bridge structure has been developed~\cite{hansen:2023} acting as virtual counterpart of a real structure present at the campus of the University of the Bundeswehr Munich, Fig.~\ref{fig:global_view}. This bridge is used as a full-scale benchmark for several studies including experimental campaign of data acquisition related to ambient and controlled damage conditions~\cite{jaelani:2023} and to testing new data acquisition strategy, e.g., the off-the-shelf use of factory smartphone accelerometers to assess structural integrity~\cite{benndorf:2016}.

\begin{figure}
    \centering
    \begin{subfigure}{0.495\textwidth}
        \centering
        \includegraphics[width=\textwidth]{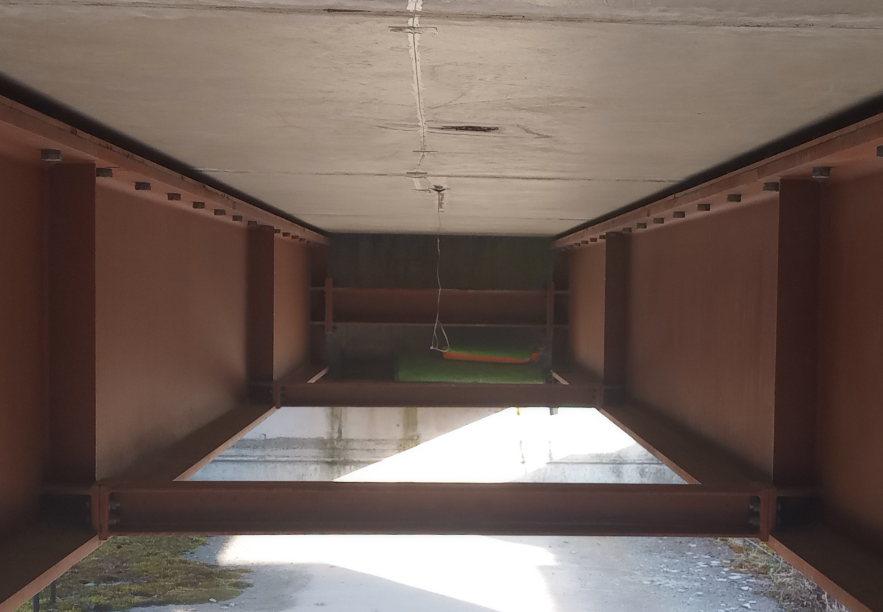}
        \subcaption{}
        \label{subfig:view_under_real}
    \end{subfigure}
    \hfill
    \begin{subfigure}{0.495\textwidth}
        \centering
        \includegraphics[width=\textwidth]{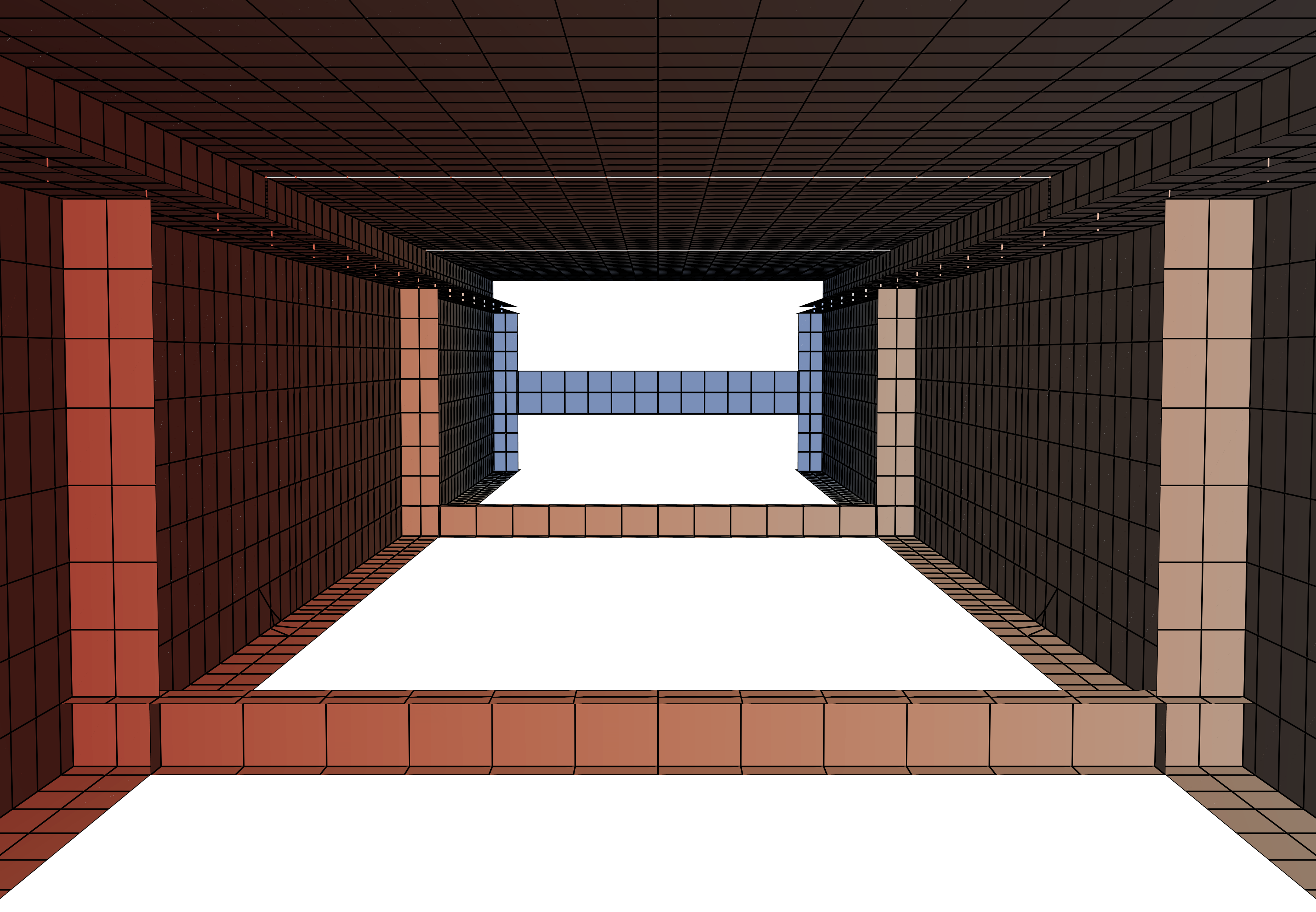}
    \subcaption{}
        \label{subfig:view_under_model}
    \end{subfigure}
    \\
    \begin{subfigure}{0.495\textwidth}
        \centering
        \includegraphics[width=\textwidth]{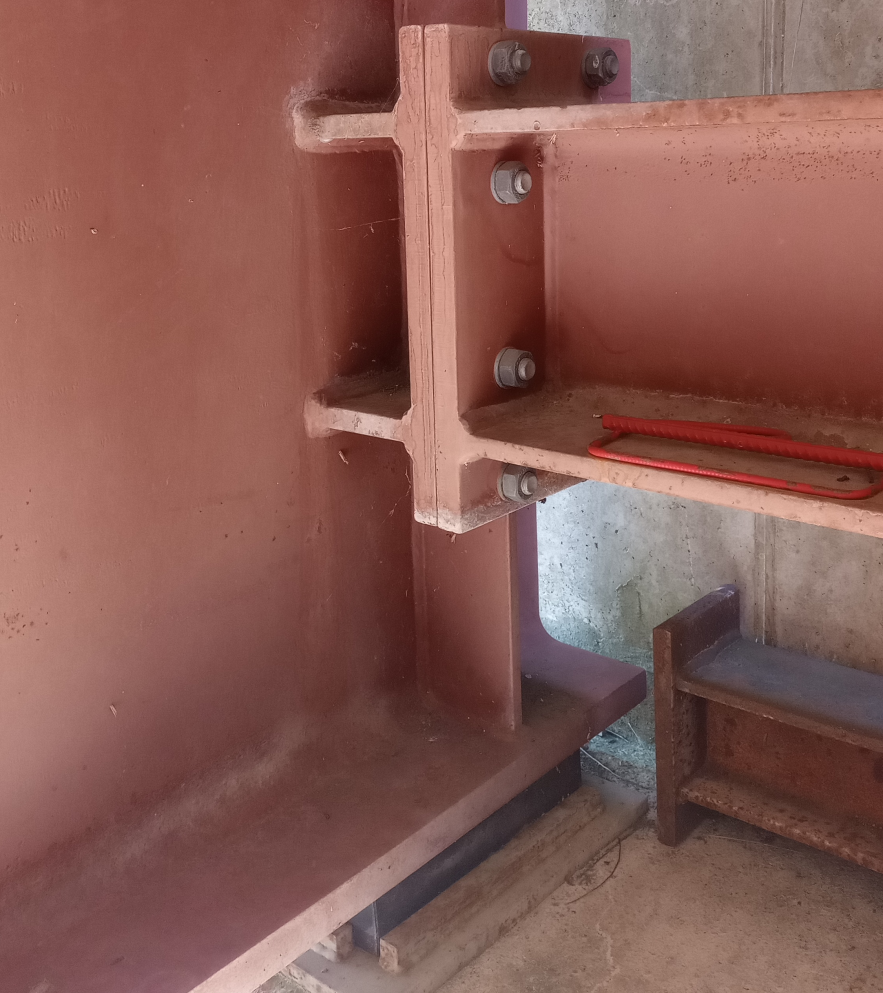}
        \subcaption{}
        \label{subfig:detail_real}
    \end{subfigure}
    \hfill
    \begin{subfigure}{0.495\textwidth}
        \centering
        \includegraphics[width=\textwidth]{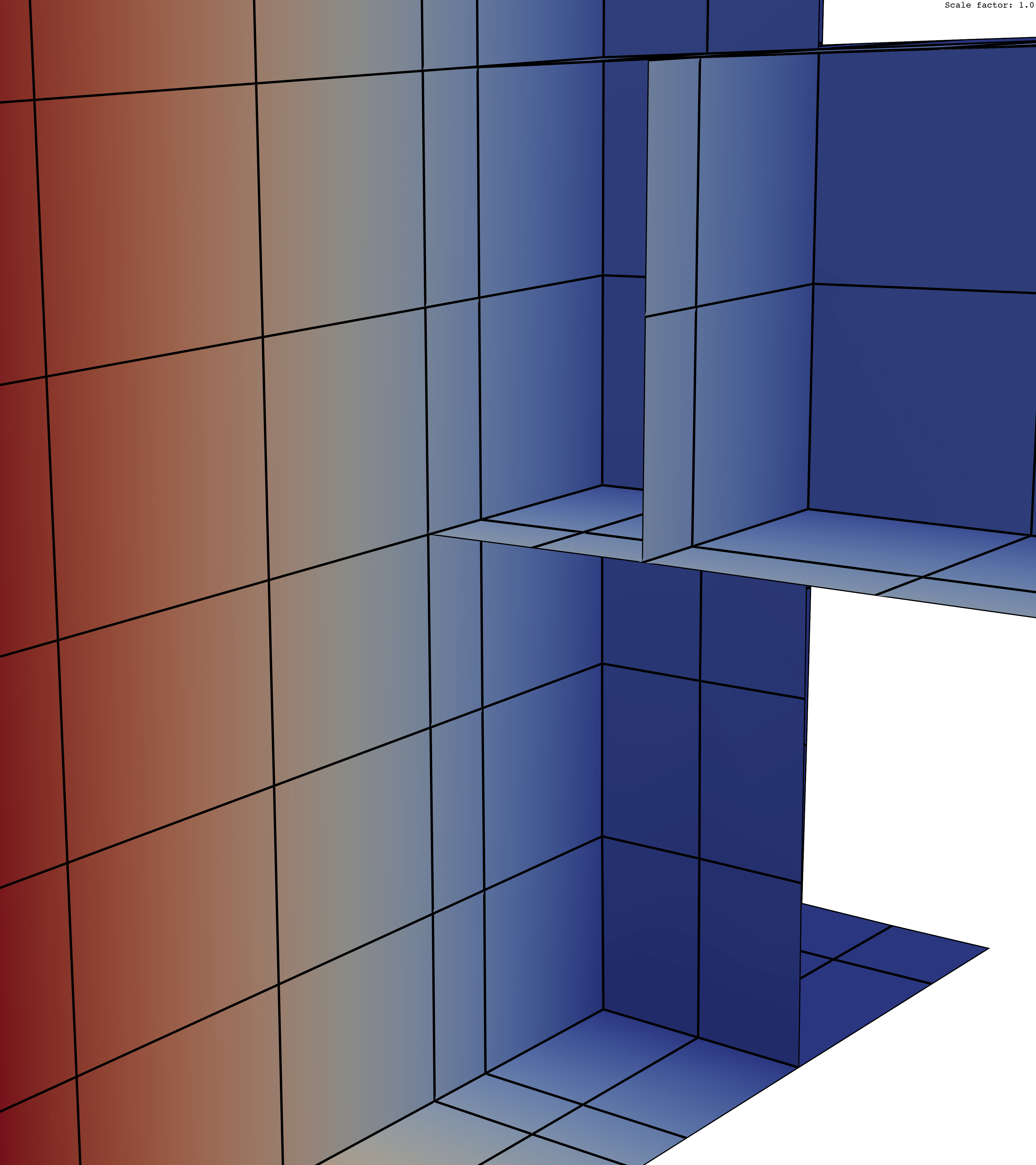}
        \subcaption{}
        \label{subfig:detail_model}
    \end{subfigure}
    \caption{Different views of the real structure compared to a realization of the FE model. Figure~(\subref{subfig:view_under_real}) features a view of the structure as seen from just below the deck, where the bolted connection between the concrete slab segments and the main HEB1000 girders is visible; Figure~(\subref{subfig:detail_real}) shows a detail of the flanged connection between one of the two main girders and the HEB280 profile stiffener located in correspondence of the eastern abutment. In parallel, Figs.~(\subref{subfig:view_under_model}) and~(\subref{subfig:detail_model}) show the equivalent parts in the virtual domain.} 
    \label{fig:struct_det}
\end{figure}

The structure is a $30~\si{\meter}$ long, two-span mixed steel-concrete bridge with two main girders and a segmented deck. The related FE model acts as a digital shadow of the real structure, and has originally been coded in \texttt{ANSYS APDL}. The model seeks a trade-off between accuracy and efficiency. Shell elements have been used consistently to model the segmented concrete deck and the principal components, i.e., webs, flanges, and stiffening plates, of all the steel members, which are HEB1000 profiles for the main girders and HEB280 and HEB120 profiles for the transverse braces. Beam elements have been used to model the bolts that connect the concrete deck to the main girders, while distributed springs connect the concrete segments to each other, to guarantee the transfer of normal stresses at the deck level. A close-up view of the real structure, together with its virtual counterpart can be observed in Fig.~\ref{fig:struct_det}.

In a subsequent design stage, the same model has been ported to \texttt{PyAnsys\footnote{Python interface to \texttt{ANSYS}; full documentation at: \url{https://docs.pyansys.com/}.}}, to guarantee an integration with state-of-the-art data science workflows and full automation of virtual SHM pipelines, without solution of continuity.

\section{gPCE-Based Sensitivity Analysis}
To evaluate the influence of input uncertainties on the model output, and gain a deeper understanding of the most relevant and influential modeling aspects, a SA study is performed using the gPCE method~\cite{sudret:2008}. This approach leverages a closed form surrogation of the model to compute the Sobol' indices, the set of parameters that describe how much the variance of several inputs of a model contributes to the global variance of a specified output~\cite{sobol:2001}. By using this approach, the indices can be evaluated analytically, thus reducing the problem complexity to the evaluation of the polynomial coefficients in a finite series expansion. To better understand the structural behavior of the test-bridge, the FE model is leveraged, and some specific inputs, judged critical for their alleged influence on the response, identified. These are: 

\begin{itemize}
\item The Young's modulus $\Ec$ of the concrete segments, a quantity for which no accurate knowledge is available, but only rough estimates.
\item The stiffness $\keq$ of the distributed springs employed to model the transfer of normal stress across the concrete segments, this uncertainty stemming from lack of informed knowledge on the material employed to realize the infill between the segments.
\item The stiffness of the bolted connection, expressed in terms of the diameter of the screw stem $\db$. While it can be argued that an accurate value of this quantity can be obtained fairly easily, considering this parameter uncertain allows, nevertheless, to assess how much the chosen modeling approach is relevant for the output under consideration, to point the designer towards more accurate modeling approaches.
\end{itemize}
As model output, four selected modes of vibration are chosen and described by the respective eigenfrequencies $f_i$. The modes analyzed are:
\begin{itemize}
\item Principal in-plane bending mode, mode $1$, $f_1\,(\si{\hertz})$.
\item First out-of-plane bending mode, mode $5$, $f_5\,(\si{\hertz})$.
\item Second out-of-plane bending mode, mode $10$, $f_{10}\,(\si{\hertz})$.
\item Principal torsional mode, mode $6$, $f_6\,(\si{\hertz})$.
\end{itemize}

\subsection{Generalized Polynomial Chaos Expansion}
Given a stochastic model $\mathcal{M}(\bX)$, function of a random variable (RV), vector input $\bX = [X_1, X_2, \ldots, X_d]^\intercal$, the Hoeffding-Sobol' decomposition~\cite{sobol:2001} allows to write it as a series made of as many terms as each possible combination of the input vector elements:
\begin{equation*}
\mathcal{M}(\bX) = \mathcal{M}_0+\sum_{i=1}^d \mathcal{M}_i(X_i) + \sum_{1\le i<j\le d}\mathcal{M}_{ij}(X_{ij})+\cdots+\mathcal{M}_{1,2,\cdots,d}(X_i,\dots,X_d).
\label{eq:hs}
\end{equation*}
In turn, the gPCE method allows to approximate the scalar model output $ Y $ in the form of a truncated polynomial expansion:
\begin{equation}
Y = \mathcal{M}(\bX) \approx \tilde{\mathcal{M}}(\bX) = \sum_{\bal \in \mathbb{N}^d} y_{\alpha} \Psi_{\bal}(\bX),
\label{eq:hs_poly}
\end{equation}
in which $ y_{\alpha} $ are the expansion coefficients, and $ \Psi_{\alpha}(\bX) $ are multivariate polynomials orthogonal with respect to the joint multivariate probability distribution that best describes $ \bX $. The \emph{multi-index} $ \bal = [\alpha_1, \alpha_2, \ldots, \alpha_d]^\intercal $ is a measure of the polynomial degree across each dimension, with a maximum value $p$ that follows:
\begin{equation}
p = \lvert \bal \rvert\ = \sum_{i=1}^d \alpha_i.
\end{equation}
The polynomials \( \Psi_{\bal}(\bX) \) form a basis of the probability space named \emph{polynomial chaos}~\cite{wiener:1938} and are constructed as the product of the univariate polynomials:
\begin{equation}
\Psi_{\bal}(\bX) = \prod_{i=1}^d P_{\alpha_i}^{(i)}(X_i),
\end{equation}
where $ P_{\alpha_i}^{(i)}(X_i) $ is the $ \alpha_i^{\text{th}} $ degree polynomial orthogonal to the respective probability density functions of $ X_i $. The choice of the polynomials to be employed is strictly dependent on the PDF describing each RV $X_i$, and different families of polynomials can be chosen accordingly, in compliance with the \emph{Askey} scheme~\cite{askey:1985}. If the joint multivariate PDF of $\bX$ is introduced as:
\begin{equation}
f_{\bX}(\bx) = \prod_{i=1}^d f_{X_i}(x_i),
\label{eq:joint}
\end{equation}
then, for each marginal distribution $f_{X_i}(x_i)$, a family of orthogonal polynomials can be defined, such that:
\begin{equation}
\langle P_j^{(i)}(x_i),P_k^{(i)}(x_i)\rangle =
\int_{\domain_i} P_j^{(i)}(x_i)P_k^{(i)}f_{X_i}(x_i)\,\mathrm{d}x_i = \gamma_j^{(i)}\delta_{jk},\label{eq:orthonorm}
\end{equation}
where $\gamma_j^{(i)}$ is a normalization constant, $\delta_{jk}$ is the Kronecker Delta and $\mathcal{D}_i$ is the integration domain. In this study, Hermite polynomials will be employed under the assumption of normally distributed inputs. For the application of the method, it is of utmost importance to notice how Eq.~\eqref{eq:orthonorm} is coincident with the expectation of the product of the polynomials. By definition of expected value, it then holds:
\begin{equation}
\mathbb E[P_j^{(i)}(X_i)P_k^{(i)}(X_i)] = \gamma_j^{(i)}\delta_{jk}.
\end{equation}
This properties guarantees that the Sobol' indices can be evaluated instantly by a proper summation of the series coefficients $y_\alpha$ alone, as explained in the next paragraphs.

It has to be noted how the orthogonal polynomials of the Askey scheme are only defined for \emph{reduced} variables of \emph{classical families} of PDF. In practical problems, this is usually not the case. Often, an additional step has to be performed to reduce the variables to standard form through a transform, that takes the general form:
\begin{equation}
X_i = F_{X_i}^{-1}\bigl(\Phi(\xi_i)\bigr),
\label{eq:map}
\end{equation}
and can be employed to map a RV $X_i$ characterized by a certain PDF, to a new standard RV. For example, a log-normal RV $ X_i\sim \mathcal{L}(\mu,\sigma^2)$ can be transformed to a new standard RV $\xi_i\sim\mathcal{N}(0,1)$. For a detailed list of classic families of orthogonal polynomials and their derivation the reader is referred to~\cite{wiener:1938}. All the steps necessary to perform this kind of analysis are exposed in detail in~\cite{sudret:2021}.

\subsection{Derivation of Sobol' indices}

Sobol' indices are variance-based sensitivity measures that decompose the output variance into contributions from individual input variables, together with the contribution given by their interactions. Starting from Eq.~\eqref{eq:hs}, the total variance $D$ of the output can be decomposed as:
\begin{equation}
D = \sum_{i=1}^d D_i + \sum_{1\le i<j\le d}D_{ij}+\cdots+D_{1,2,\cdots,d}.
\label{eq:dec}
\end{equation}
By definition, the Sobol' indices can be then introduced dividing both members of Eq.~\eqref{eq:dec} by the total variance, delivering:
\begin{equation}
1 = \sum_{i=1}^d S_i + \sum_{1\le i<j\le d}S_{ij}+\cdots+S_{1,2,\cdots,d}.
\label{eq:sob}
\end{equation}
Each index $S_{i_1,\cdots,i_s}$ that appears on the right-hand side of Eq.~\eqref{eq:sob} is a sensitivity measure that describes which fraction of $D$ is due to the variance of each input parameter $X_i$. The indices can be classified in first order indices, in number equal to $d$, and higher order indices, that describe the combined influence of more input parameters acting together. The overall number of indices for a model with $d$ input parameter is $2^d-1$. Sobol' indices can be evaluated in a standard way using Monte Carlo methods, but at the risk of a possibly unbearable computational cost.

An alternative consists in the evaluation of the Sobol' indices considering the polynomial expansion $\tilde{\mathcal{M}}(\bX)$. If it is substituted in the original Hoeffding-Sobol' model decomposition, see Eq.~\eqref{eq:hs_poly}, the orthogonality of the polynomials allows to write every index as a combination of the coefficients of the expansion alone, as:
\begin{equation}
S_{\bu} = \sum_{\bal\in\mathcal{A}_{\bu}} y_{\alpha}^2\Bigl/
\sum_{\bal\in\mathcal{A}\setminus0} y_{\bal}^2,
\label{eq:sobol_indices}
\end{equation}
where $\mathcal{A}_{\bu} = \{\bal\in\mathcal{A}:k\in\bu\iff\alpha_k \ne0\}$ excludes from the derivation the constant term of the expansion and the denominator coincides with the total variance $D$ of the system. By computing these indices, the most influential input variables and their interactions can be identified providing valuable insights into the model behavior and guiding further analyses or model improvements.

\section{Application}\label{sec:application}
The aforementioned derivation is employed to quantify the influence of selected uncertain parameters on the structural response of the test-bridge model. The analysis focuses on three critical input variables: \emph{(i)} the stiffness of the concrete composing the top deck, expressed in terms of its Young's modulus $\Ec$; \emph{(ii)} the stiffness of the bolts connecting the steel girders to the concrete slab, which affects shear transfer and composite section behavior, and is modeled by an uncertain diameter of the screw stem $\db$; the stiffness $\keq$ of the distributed springs employed to model the connection between adjacent concrete slab segments, which influences the continuity and the global flexibility of the deck. By systematically varying these parameters within plausible ranges, their individual and combined contributions to form the model output can be assessed, providing insights into the most significant contributors to variability of the bridge’s modal response, expressed by the selected modes of vibration.

The PDFs chosen to describe the three random parameters are all log-normal and fully characterized in Tab.~\ref{tab:pdfs} in terms of the two parameters $\mu$ and $COV = \sigma/\mu$. From now on, to maintain consistency with standard notation for stochastic quantities, the model inputs will be collected in the input vector $\bX \in \mathbb{R}^{3}$, and the eigenfrequencies in the model output $\bY \in \mathbb{R}^{4}$.\footnote{This helps distinguishing between a random variable $X_i$ and its respective physical parameters. The ordering is maintained, so that $[X_1,X_2X_3] \leftrightarrow [\Ec,\db,\keq]$ and $[Y_1,\cdots,Y_{4}] \leftrightarrow [f_1,f_5,f_6,f_{10}]$. Where not source of ambiguity, and without loss of generalization, the scalar RV will be used without its respective subscript.}

The analysis consists in two main steps, namely, the identification of a suitable polynomial basis and the identification of the series coefficients. Once this quantities have been derived, the evaluation of the Sobol' indices is straightforward, in accordance with Eq.~\eqref{eq:sobol_indices}.

\begin{table}[t]
\centering
\caption{Random input characterization. A lognormal distribution is assumed for each of the three parameters concrete stiffness $\Ec$, connecting bolt screw stem $\db$, and deck connection equivalent spring stiffness $\keq$.}
\label{tab:pdfs}
\begin{tabular}{cccr}
\toprule
RV & Parameter & $\mu$ & $COV$\\
\midrule
$X_1$ & $\Ec$ & $35.0~\si{\giga\pascal}$ & $5.7~\%$ \\
$X_2$ & $\db$ & $18.0~\si{\milli\meter}$ & $11.1~\%$ \\
$X_3$ & $\keq$ & $30.1~\si{\mega\newton/\milli\meter}$ & $11.2~\%$ \\
\bottomrule
\end{tabular}
\end{table}

\subsection{Polynomial basis identification}
To construct the polynomial expansion, the Python library \texttt{Chaospy}~\cite{feinberg:2015} has been employed. As a first step, the polynomial basis for the expansion is constructed, fully defined by the joint probability distribution $f_{\bX}(\bx)$, cf. Eq.~\eqref{eq:joint}, that dictates the type of polynomials to be employed, and by a chosen maximum degree $p$. Since all the univariate probability distributions are log-normal, all the univariate polynomials composing each multivariate term of the expansion will be Hermite polynomials, $\he(\cdot)$. As a preparatory step, two transformations must be applied to guarantee the orthogonality of the polynomials. The first maps the log-normal distribution $F(x_i)$ to a gaussian distribution $\Phi(\xi_i)$, cf. Eq.~\eqref{eq:map}, the second is a isoprobabilistic transform that enforces $\xi_i \sim \mathcal{N}(0,1)$. While these two steps could be conveniently condensed in a single operation, the second one is automatically managed by \texttt{Chaospy} when defining the joint distribution, so that only the first must be explicitly performed by the user. With this information, the polynomial expansion can be built. Out of completeness, an example of the multi index $\bal$ for a dummy case $p=2$ is exposed in Tab.~\ref{tab:mi}, together with the multivariate polynomials composing each term of the summation. For this chosen $p$ and three inputs, the summation accounts for a total number of terms $P$ equal to ten. In the application addressed, both three- and four-order degree expansions have been emplyed.

\begin{table}[t]
\centering
\caption{Multi indices $\bal$ for an expansion of degree $p=2$ and $d=3$ inputs.}
\label{tab:mi}
\begin{tabular}{cccr}
\toprule
Index & $\bal$ & $\Psi_{\bal}(\bxi)$ \\
\midrule
$0$ & $[0,0,0]$ & $\he_0(\xi_0)\cdot\he_0(\xi_1)\cdot\he_0(\xi_2)$ \\
$1$ & $[0,0,1]$ & $\he_0(\xi_0)\cdot\he_0(\xi_1)\cdot\he_1(\xi_2)$ \\
$2$ & $[0,0,2]$ & $\he_0(\xi_0)\cdot\he_0(\xi_1)\cdot\he_2(\xi_2)$ \\
$3$ & $[0,1,0]$ & $\he_0(\xi_0)\cdot\he_1(\xi_1)\cdot\he_0(\xi_2)$ \\
$4$ & $[0,1,1]$ & $\he_0(\xi_0)\cdot\he_1(\xi_1)\cdot\he_1(\xi_2)$ \\
$5$ & $[0,2,0]$ & $\he_0(\xi_0)\cdot\he_2(\xi_1)\cdot\he_0(\xi_2)$ \\
$6$ & $[1,0,0]$ & $\he_1(\xi_0)\cdot\he_0(\xi_1)\cdot\he_0(\xi_2)$ \\
$7$ & $[1,0,1]$ & $\he_1(\xi_0)\cdot\he_0(\xi_1)\cdot\he_1(\xi_2)$ \\
$8$ & $[1,1,0]$ & $\he_1(\xi_0)\cdot\he_1(\xi_1)\cdot\he_0(\xi_2)$ \\
$9$ & $[2,0,0]$ & $\he_2(\xi_0)\cdot\he_0(\xi_1)\cdot\he_0(\xi_2)$ \\
\bottomrule
\end{tabular}
\end{table}

\subsection{Series coefficients evaluations}
A non-intrusive regression method has been used to evaluate the coefficients of the series. The coefficients vector is obtained minimizing the mean-squared residual error:
\begin{equation}
\hat{\boldsymbol{y}} = \arg\min_{\boldsymbol{y}}\mathbb{E}\biggl[~\sum_{\bal\in\mathcal{A}}y_\alpha\Psi_{\bal}-\mathcal{M}(\boldsymbol{\mathcal{X}})\biggl],
\end{equation}
where $\hat{\boldsymbol{y}}$ collects the $P$ series coefficients and $\boldsymbol{\mathcal{X}}=[\bx^{(1)},\dots,\bx^{(n)}]^\intercal$ represents an experimental design, i.e., a specified sampling of the inputs that best covers the parameters space; $\mathcal{M}(\boldsymbol{\mathcal{X}})$ is then defined by running the high-fidelity model on the experimental design in a Monte Carlo fashion. An empirical rule suggests the experimental design size $n=(d-1)P$~\cite{sudret:2008}. During the solution stage, it has been verified that increasing the size of the experimental design above $n=10$ did not bring any improvement in the accuracy of the results, neither for $p=3$ nor $p=4$. In the sampling process from the input parameter space, a random method to choose the points has been used, with a fixed seed to guarantee reproducibility of the results.

\section{Results and Discussion}
\begin{table}[t]
\centering
\caption{Look-up table for global sensitivity analysis results in terms of Sobol' indices, negligible terms are grayed out and different order indices are separated by a horizontal rule.}
\label{tab:sensitivity_complete}
\begin{tabular}{l cccccccc}
\toprule
\multirow{2}*{Index}
& \multicolumn{2}{c}{Mode 1} & \multicolumn{2}{c}{Mode 5}
& \multicolumn{2}{c}{Mode 6} & \multicolumn{2}{c}{Mode 10}\\
\cmidrule(lr){2-3} \cmidrule(lr){4-5} \cmidrule(lr){6-7} \cmidrule(lr){8-9}
& $p=3$ & $p=4$ & $p=3$ & $p=4$ & $p=3$ & $p=4$ & $p=3$ & $p=4$ \\
\midrule
$S_1$     & 0.15650 & 0.15660 & 0.01443 & 0.01445 & 0.63487 & 0.63516 & 0.23720 & 0.23740 \\
$S_2$     & 0.84349 & 0.84339 & 0.98550 & 0.98549 & 0.36506 & 0.36477 & 0.76275 & 0.76256 \\
$S_3$     & \gr{0.00000} & \gr{0.00000} & \gr{0.00000} & \gr{0.00000} & \gr{0.00000} & \gr{0.00000} & \gr{0.00000} & \gr{0.00000} \\
\midrule
$S_{12}$  & 0.00002 & 0.00002 & 0.00007 & 0.00006 & 0.00007 & 0.00007 & 0.00004 & 0.00004 \\
$S_{23}$  & \gr{0.00000} & \gr{0.00000} & \gr{0.00000} & \gr{0.00000} & \gr{0.00000} & \gr{0.00000} & \gr{0.00000} & \gr{0.00000} \\
$S_{13}$  & \gr{0.00000} & \gr{0.00000} & \gr{0.00000} & \gr{0.00000} & \gr{0.00000} & \gr{0.00000} & \gr{0.00000} & \gr{0.00000} \\
$S_{123}$ & \gr{0.00000} & \gr{0.00000} & \gr{0.00000} & \gr{0.00000} & \gr{0.00000} & \gr{0.00000} & \gr{0.00000} & \gr{0.00000} \\
\midrule
$S_{T1}$  & 0.15651 & 0.15661 & 0.01450 & 0.01451 & 0.63494 & 0.63523 & 0.23725 & 0.23744 \\
$S_{T2}$  & 0.84350 & 0.84340 & 0.98557 & 0.98555 & 0.36513 & 0.36484 & 0.76280 & 0.76260 \\
$S_{T3}$  & \gr{0.00000} & \gr{0.00000} & \gr{0.00000} & \gr{0.00000} & \gr{0.00000} & \gr{0.00000} & \gr{0.00000} & \gr{0.00000} \\
\bottomrule
\end{tabular}
\end{table}

\begin{figure}
    \centering
    \begin{subfigure}{0.495\textwidth}
       \centering
        \includegraphics[width=\textwidth]{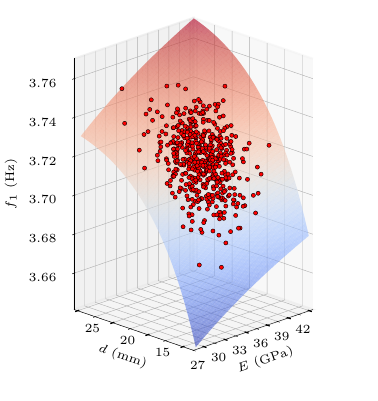}
        \subcaption{}
        \label{subfig:surf_1}
    \end{subfigure}
    \begin{subfigure}{0.495\textwidth}
        \centering
        \includegraphics[width=\textwidth]{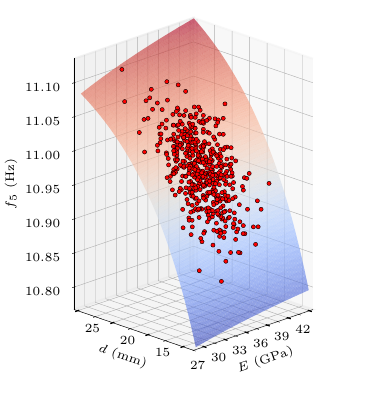}
        \subcaption{}
        \label{subfig:surf_5}
    \end{subfigure}\\
    \begin{subfigure}{0.495\textwidth}
       \centering
        \includegraphics[width=\textwidth]{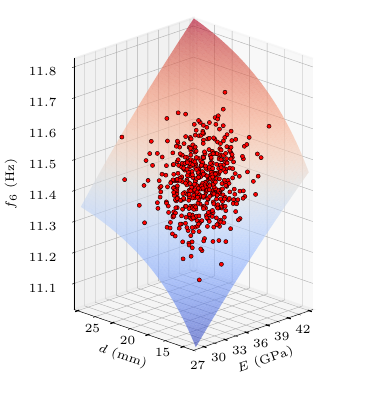}
        \subcaption{}
        \label{subfig:surf_6}
    \end{subfigure}
    \begin{subfigure}{0.495\textwidth}
        \centering
        \includegraphics[width=\textwidth]{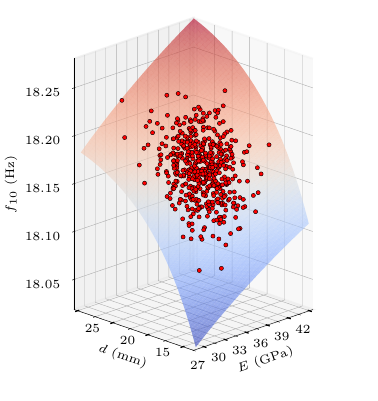}
        \subcaption{}
        \label{subfig:surf_10}
    \end{subfigure}
    \caption{Response surfaces for the eigenfrequencies plotted in function of $\Ec$ and $\db$. All the models show an almost complete (in)sensitivity with respect to $\keq$, which is, therefore, not shown. Three modes of vibration manifest a higher sensitivity with respect to $\db$ compared to $\Ec$,~(\subref{subfig:surf_1}),~(\subref{subfig:surf_5}), and~(\subref{subfig:surf_10}), except for the torsional mode,~(\subref{subfig:surf_6}), which is more sensitive to the concrete stiffness. All the frequency scales are expressed in $(\si{\hertz})$.}
    \label{fig:response_surfaces}
\end{figure}

\noindent
After the complete determination of the surrogate model $\tilde{\mathcal{M}}(\bx)$, the Sobol' indices, all first order, higher order and total order, can finally be evaluated according to Eq.~\eqref{eq:sobol_indices}. The results in terms of response surfaces are plotted in Fig.~\ref{fig:response_surfaces} for the selected proper modes of vibration of the structure, correspondent to the principal in-plane vibration mode, mode $1$, principal out-of-plane bending mode, mode $5$, principal torsional mode, mode $6$ and, finally, secondary out-of-plane bending mode, mode $10$. Each response surface is plotted together with a realization of the experimental design $\boldsymbol{\mathcal{X}}$, superposed as a scatter plot with red circular markers. The surrogate models represented are related to a polynomial expansion with $p=4$. It can be observed that all the models feature a marked dependency from the stiffness of the bolts, parametrized by $\db$, represented on the first coordinate axis, together with an appreciable dependency on the stiffness of the concrete slab $\Ec$, except for mode $5$, which is almost completely unsensitive with respect to a variation in concrete rigidity. A common feature for all the four models is the almost complete independence from the stiffness of the concrete joints.

\begin{landscape}
\begin{figure}[htbp]
    \centering
    \begin{subfigure}{0.495\linewidth}
       \centering
        \includegraphics[width=\textwidth]{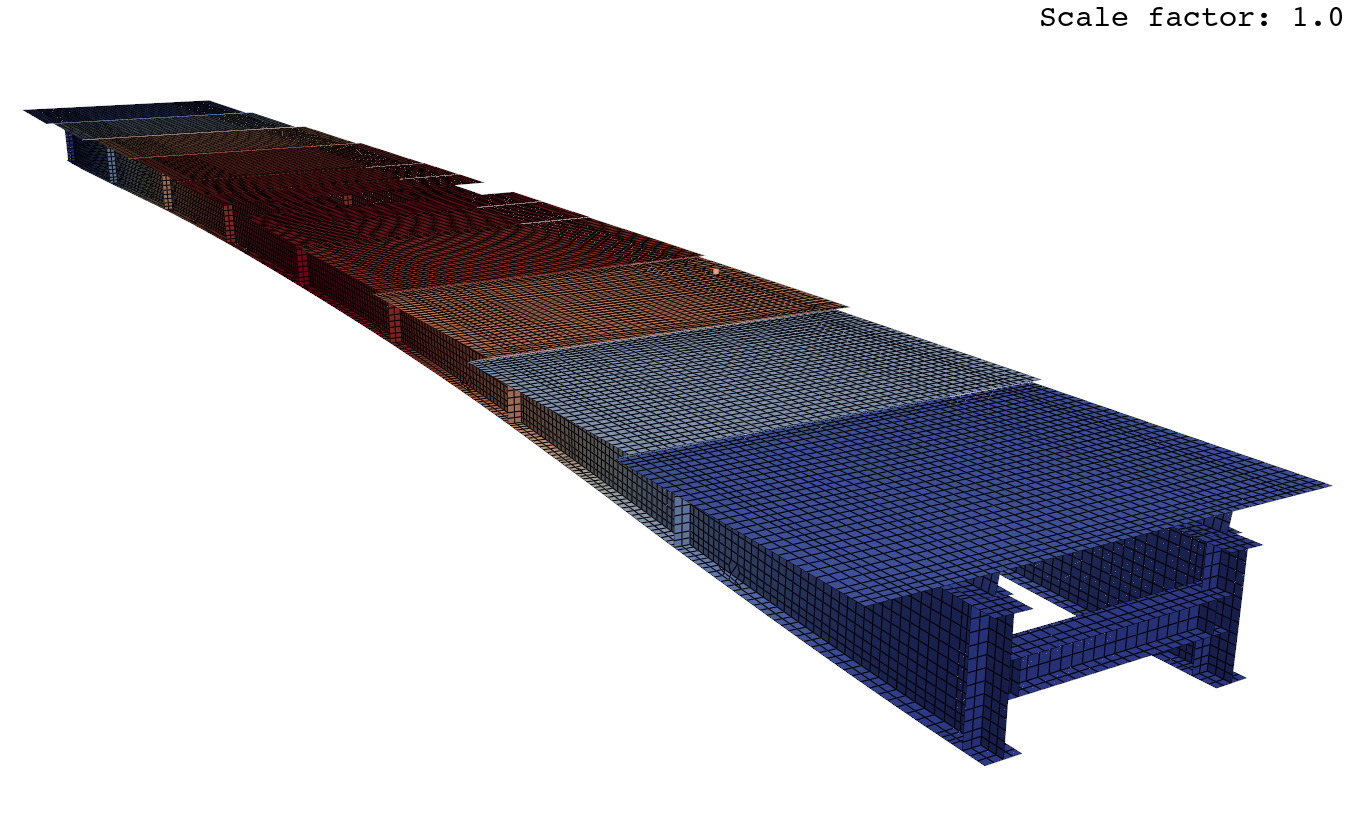}
        \subcaption{}
        \label{subfig:mode1}
    \end{subfigure}
    \begin{subfigure}{0.495\linewidth}
        \centering
        \includegraphics[width=\textwidth]{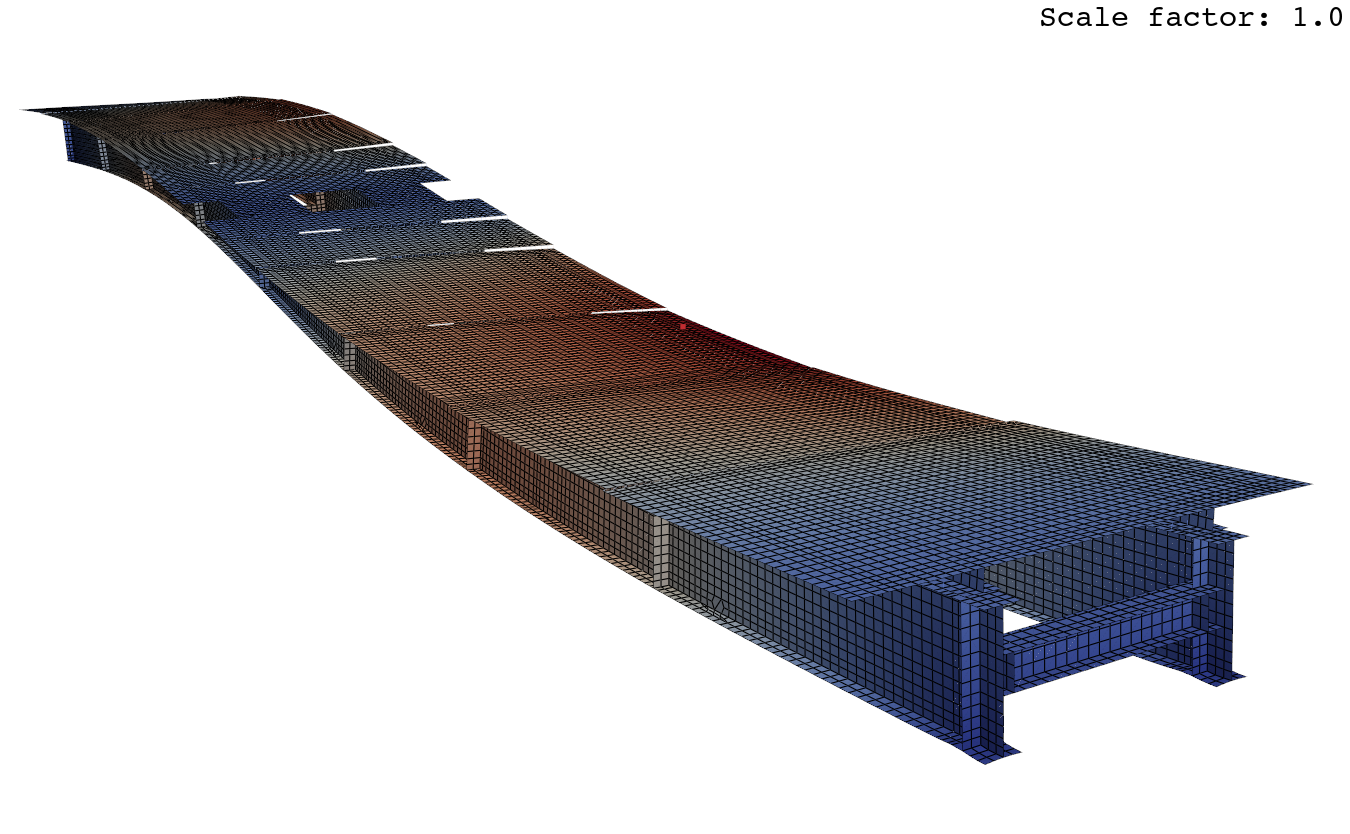}
        \subcaption{}
        \label{subfig:mode5}
    \end{subfigure}\\
    \begin{subfigure}{0.495\linewidth}
       \centering
        \includegraphics[width=\textwidth]{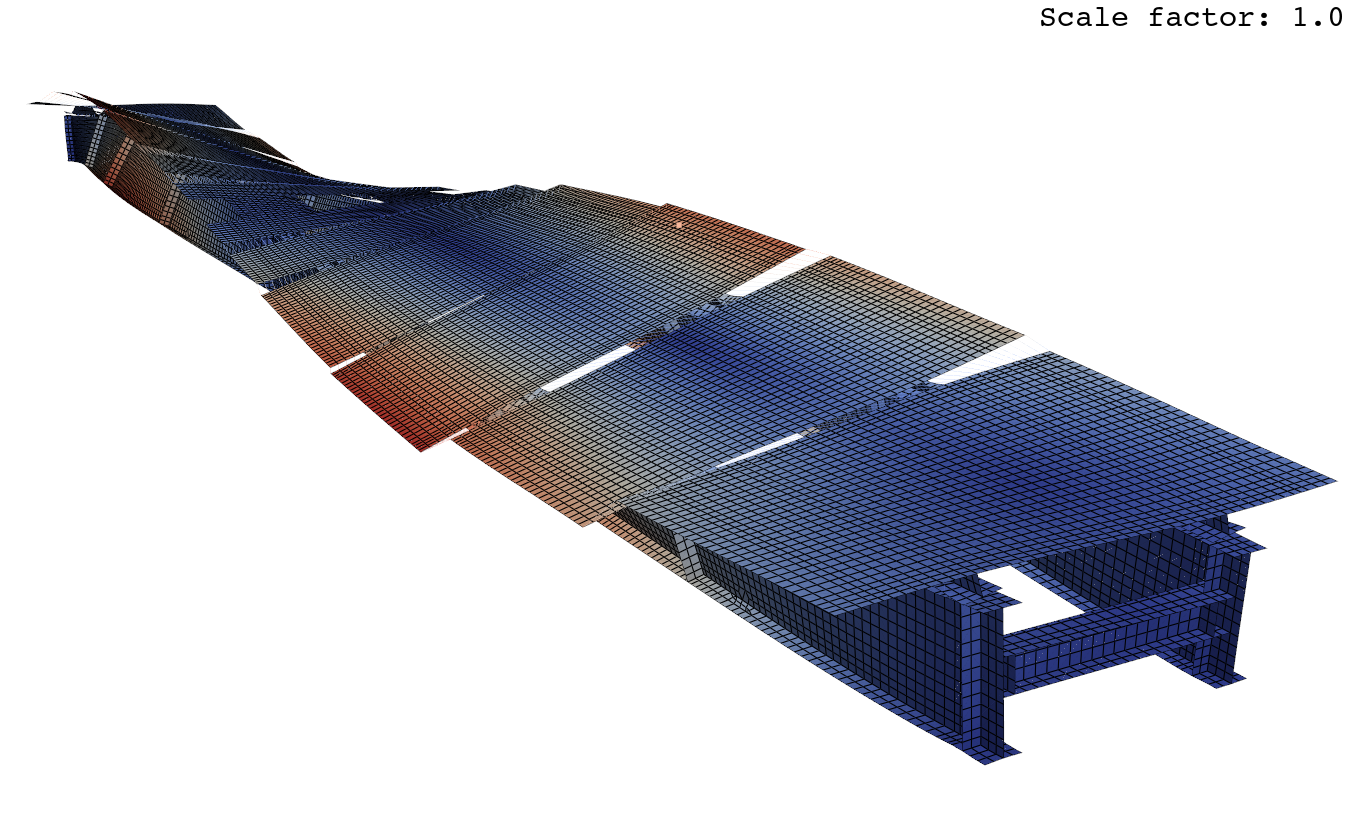}
        \subcaption{}
        \label{subfig:mode6}
    \end{subfigure}
    \begin{subfigure}{0.495\linewidth}
        \centering
        \includegraphics[width=\textwidth]{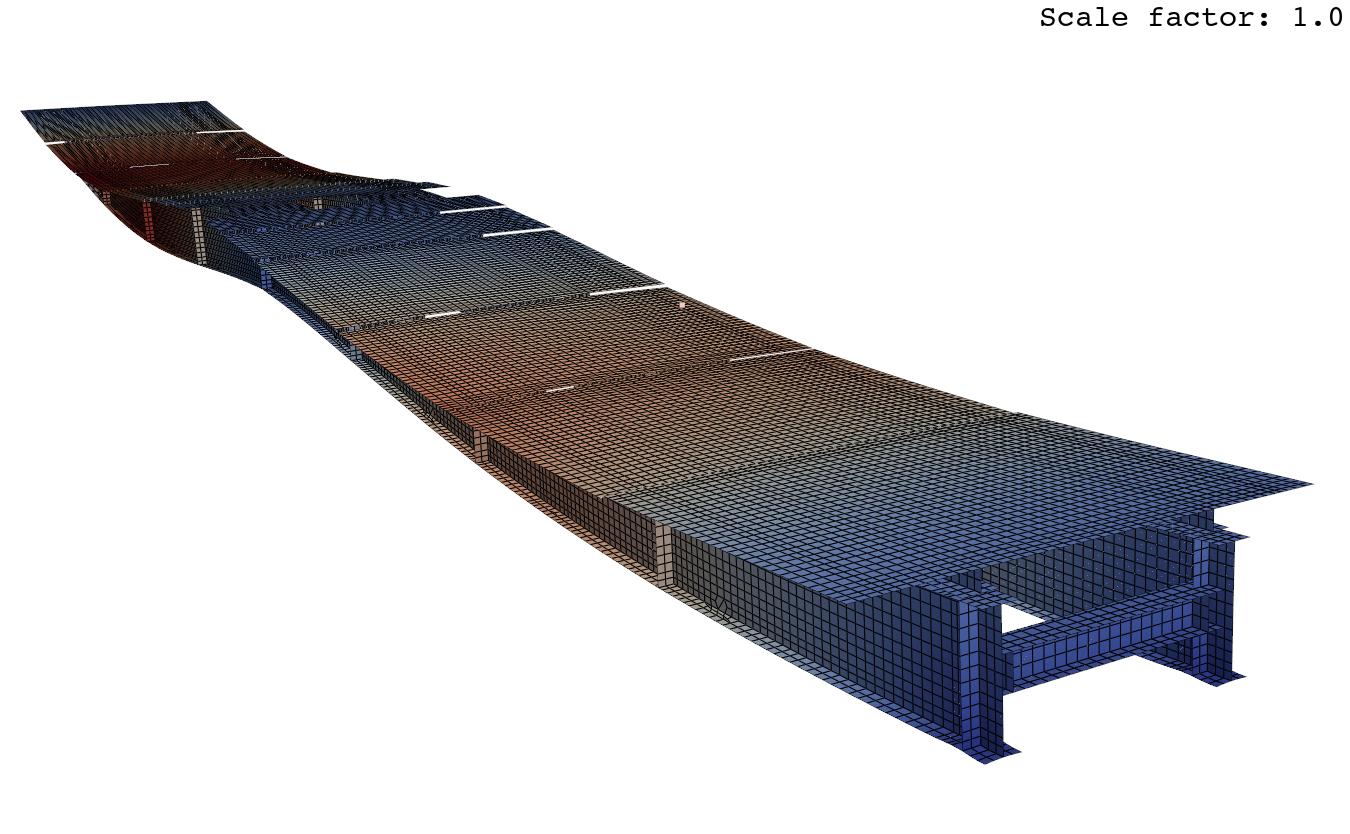}
        \subcaption{}
        \label{subfig:mode10}
    \end{subfigure}
    \caption{Results of the FE model, for a given value of the experimental design, for the selected modes of vibration analyzed, i.e., in-plane bending~(\subref{subfig:mode1}); principal out-of-plane bending~(\subref{subfig:mode5}); principal torsion~(\subref{subfig:mode6}); secondary out-of-plane bending~(\subref{subfig:mode10}).}
    \label{fig:final_model}
\end{figure}
\end{landscape}

A quantitative evaluation of the sensitivity can be made by analyzing the Sobol' indices, which are reported in Tab.~\ref{tab:sensitivity_complete} in their full set, i.e., first order, second order, higher order, and total order, the last being obtained by summing all the indices containing a selected output. A general trend can be observed in the identification of the $\db$ as the parameter the model is more sensitive to, except for the torsional mode, for which the stiffness of the concrete is remarkably more influential. For all the vibration modes taken into consideration, the stiffness of the connection between the concrete slabs does not appear to be important, leading to negligible values of the indices, regardless their order. If, from one hand, this could stem from an inaccurate an perhaps too simplistic modeling of the connection, an effect should at least be observed in the modes of vibration experiencing out of plane bending, i.e., modes $5$ and $10$. This two modes operate a transmission of force inherently driven by normal stress, compliant with the type of spring elements employed at the modeling stage.
Finally, global views of the four mode shapes related to the modes of vibration analyzed are shown in Fig.~\ref{fig:final_model}, that showcases the dynamic behavior of the structure.

\section{Conclusions and Future Perspectives}

The study presented a SA framework applied to the FE model of a full-scale test bridge, combining the high-fidelity proper of a physics-based model with an advanced uncertainty quantification technique. By dealing with a virtual replica and applying the gPCE method, the most influential of the selected structural parameters affecting the bridge's dynamic response were successfully identified. The results demonstrate that while material properties and connection details significantly impact vibration characteristics, certain parameters exhibit negligible influence, offering valuable insights for model simplification, possible model improvements, and targeted monitoring strategies.

The integration of Sobol' sensitivity indices with state-of-the-art computational workflows provides a robust foundation for physics-informed decision-making in infrastructure management. This approach not only enhances the understanding of structural behavior under uncertainty, but also contributes to the development of more efficient monitoring systems for critical infrastructures. Furthermore, looking beyond individual bridge applications, the proposed framework could be extended to population-based analyses of similar structures, offering substantial benefits for a global approach to the protection of critical infrastructure networks~\cite{gardner:2021}, an objective that can be reached focusing not only on the engineering dimension, but broaden the view to account for, e.g., extended dimensions like social vulnerability~\cite{malinowska:2025}. Bringing this global view into maintenance strategies would allow for holistic infrastructure management and optimized resource allocation, thus bringing currently restricted frameworks to a broader scale, well-equipped and more suitable to face the challenges of contemporary times.

\printbibliography

\end{document}